\documentclass[nonacm]{acmart}
\AtBeginDocument{%
  }

\setcopyright{acmlicensed}
\copyrightyear{2027}
\acmYear{2027}
\acmDOI{XXXXXXX.XXXXXXX}
\acmConference[CS\&Law '27]{6th ACM Symposium on Computer Science and Law}{March 23--25, 2027}{Chicago, IL, USA}
\acmISBN{978-1-4503-XXXX-X/2018/06}

\usepackage{todonotes}
\usepackage{enumitem}
\setlist[itemize]{leftmargin=*}
\setlist[enumerate]{leftmargin=*}
\usepackage{multirow}
\usepackage{soul}
\usepackage{tabularx}
\usepackage{booktabs}
\usepackage{array}
\usepackage{subcaption}
\usepackage{longtable}
\usepackage{placeins}
\usepackage{needspace}

\newlist{questions}{enumerate}{2}
\setlist[questions,1]{label=\textbf{RQ\arabic*.},ref=RQ\arabic*, left=0em, align=left}
\setlist[questions,2]{label=(\alph*),ref=\thequestionsi(\alph*)}

\newenvironment{typology}[1]{%
  \begin{enumerate}[label=\textbf{#1\arabic*.},ref=#1\arabic*,left=0em,align=left]%
}{%
  \end{enumerate}%
}

\begin{document}

\title{Privacy Preference Controls for Agentic Systems}

\author{Shreya Kochar}
\authornote{Corresponding authors.}
\email{shkochar@ucsd.edu}
\affiliation{%
  \institution{University of California, San Diego}
  \city{San Diego}
  \state{CA}
  \country{USA}
}

\author{Peeranat Sugsavanvit}
\affiliation{%
  \institution{Wesleyan University}
  \city{Middletown}
  \state{CT}
  \country{USA}
}

\author{Vidur Gupta}
\email{vidur@stanford.edu}
\affiliation{%
  \institution{Stanford University}
  \city{Stanford}
  \state{CA}
  \country{USA}
}

\author{Joseph A. Calandrino}
\affiliation{%
  \institution{Carnegie Mellon University}
  \city{Pittsburgh}
  \state{PA}
  \country{USA}
}

\author{Deepak Kumar}
\affiliation{%
  \institution{University of California, San Diego}
  \city{San Diego}
  \state{CA}
  \country{USA}
}

\author{Brett Frischmann}
\affiliation{%
  \institution{Villanova University}
  \city{Villanova}
  \state{PA}
  \country{USA}
}

\author{Sebastian Zimmeck}
\authornotemark[1]
\email{szimmeck@wesleyan.edu}
\affiliation{%
  \institution{Wesleyan University}
  \city{Middletown}
  \state{CT}
  \country{USA}
}

\renewcommand{\shortauthors}{Kochar et al.}

\begin{abstract}
In response to vast and widespread online tracking, regulators and technologists alike have worked to develop laws and tools that allow people to express their privacy preferences. 
These measures seek to offer users greater control over the sale and sharing of personal information, putting agency in their hands to protect their privacy. 
However, the web is undergoing a seismic shift, away from strictly \textit{user-driven} behavior towards \textit{agentic behavior}, where AI agents take autonomous actions on behalf of users. 
In this study, we ask: What should privacy preference controls look like in the agentic web and agentic systems more generally? 
To answer this question, we develop a taxonomy of enforceable privacy preferences for agentic systems. 
Our taxonomy is informed both by laws and empirical studies of existing privacy tools as well as an audit of 30 leading AI assistants and agent platforms. 
To explore how our taxonomy can be implemented, we design and develop eight open-source prototypes that mirror real-world agentic applications and demonstrate how major aspects of our taxonomy can be structurally integrated into agentic protocols, such as the Model Context Protocol and Agent2Agent protocol. 
Ultimately, our study seeks to provide the groundwork for ensuring end-user privacy and autonomy are not lost in the newly arriving agentic world.
\end{abstract}


\received{20 October 2026}
\received[revised]{TBD}
\received[accepted]{TBD}

\maketitle

\section{Introduction}
\label{sec:intro}



Internet users often pay for ``free'' services with their data~\cite{hoofnagle2014free}. On the web, for example, ad networks track users via cookies, browser fingerprinting, and other privacy-invasive mechanisms~\cite{englehardtNarayanan2016}. In response to this widespread tracking, an increasing number of privacy laws, such as the California Consumer Privacy Act (CCPA)~\cite{ccpa-cpra} mandate that people have a right to opt-out from tracking. These legal requirements are often complemented by technical systems, such as Global Privacy Control (GPC)~\cite{GPCSpec}, which are intended to ease the burden of privacy self-management for users.

However, the Internet writ large is undergoing a major shift away from user-driven interactions and towards \textit{agentic} interactions~\cite{osmonbekov2026deadinternet}. In an agentic web, AI agents autonomously perform everyday tasks on behalf of users---for example, system's like Meta's Muse~\cite{MetaMuse} and OpenAI's Dots~\cite{openaiDots2026} can, without supervision, book a flight---from researching the best connections, finding the lowest price, to the actual booking~\cite{MetaMuse}. For such tasks, agents need privileged access to sensitive data, which they may overshare and use without direct user control~\cite{WSJMetaMuse, shapira2026agentschaos}. To make matters worse, AI companies have already started integrating ads to generate revenue~\cite{openai_ads2026,MicrosoftCopilotCheckout} and AI systems are freely participating in third-party tracking~\cite{oliveiraEtAlButterfly2027}.
These new, agentic modalities both fundamentally reshape our relationship with privacy and introduce a new swath of privacy challenges that we as the research community must now contend with.

In this study, we evaluate the need for and feasibility of \textit{privacy preference controls} in agentic systems.\footnote{We also refer to those as \textit{enforceable privacy preferences}, or, simply, \textit{preferences} or \textit{controls}.} To do so, we develop a taxonomy of privacy preferences that can be expressed in agentic systems, informed by three sources: (1) existing laws that enable users to express privacy preferences in current technical systems, (2) empirical opt-out studies (e.g., those studying how users engage with cookie banners), and (3) an empirical audit of AI products released by 22~companies, exploring what controls they offer to users today. We then implemented eight \textit{agentic prototypes} (Section~\ref{sec:prototypes}) informed by existing tools, to examine privacy preferences in our taxonomy. This examination demonstrates the feasibility of some existing privacy controls while also showing the need for more robust coverage.

We situate our work in stark contrast with the current practice of data maximization on the web and the \textit{surveillance capitalism} that it supports~\cite{ZuboffSurveillanceCapitalism}. Instead, we see data minimization as a legally enforceable default from which exceptions should be allowed only when demonstrably informed consent exists~\cite{frischmannRajtmajer2026}. We ground privacy preference controls in the understanding of privacy as a protective right for an individual's personal autonomy~\cite{roessler2005privacy}.
This not only encompasses which data should be shared with whom but also being free from the presence and persistence of AI agents. Our taxonomy is also informed by contextual integrity (CI), especially in understanding the boundaries of data collection and sharing in agentic systems~\cite{privacy_nissenbaum_2009, Bagdasarian2026}.
Our study seeks to answer the following questions:

\begin{questions}
    \item What privacy preference controls should people be able to express and have respected in agentic systems?
    \item What are the technological and legal challenges that need to be considered and resolved for privacy preference controls to be effective in agentic systems?
    \item How can we structurally integrate privacy preference controls into agentic systems, especially, as such systems interact with the web and online ad ecosystem?
\end{questions}

After discussing the background and related work (\S\ref{sec:background_and_related_work}) and our methodology (\S\ref{sec:methodology}), we contribute:

\begin{itemize}
    \item A taxonomy for privacy preference controls on agentic systems for implementation in agentic protocols and informed by an empirical audit (\S~\ref{sec:taxonomy}).
    \item An empirical audit of major technology and AI companies' privacy preference controls in their AI products as well as the agentic protocols they use and support (\S~\ref{sec:analysis-coverage}).
    \item Open-source prototype implementations for expressing and propagating enforceable privacy preferences for the MCP and A2A protocol (\S~\ref{sec:prototypes}).\footnote{Our prototype implementations are publicly available at \url{https://github.com/privacy-tech-lab/gpc-ai-agents-prototypes}.}
\end{itemize}

We then discuss our research questions (\S~\ref{sec:analysis_and_discussion}) in the context of broader legal concepts and the technological landscape and conclude our evaluation with future work (\S~\ref{sec:conclusions}).
\section{Background and Related Work}
\label{sec:background_and_related_work}

Our work is motivated by the failures of notice and choice (\S~\ref{sec:notic-and-choice}) and the emergence of privacy preference signals as a promising opt-out mechanism for web tracking (\S~\ref{sec:privacy-preference-signals}).
As AI agents are integrated into the online ad ecosystem (\S~\ref{sec:web-advertising-and-tracking}), we can adapt privacy preference signals for agentic protocols (\S~\ref{sec:privacy-in-agentic-protocols}) to minimize data collection and sharing and respect contextual boundaries (\S~\ref{sec:data-minimization-and-ci}).

\subsection{Notice and Choice}
\label{sec:notic-and-choice}

U.S. privacy law relies on notice and choice: organizations disclose their data practices in a privacy notice, and users make their choice, whether through explicit agreement, an objection, or continued use. 
In most other countries, opt-in consent is the norm~\cite{yu_yin_zimmeck_2026}. 
Notably, the General Data Protection Regulation (GDPR) prohibits personal data processing unless one of six lawful bases, among which opt-in consent is the most common, is present.

The shortcomings of notice and choice are well-documented.
Notice and choice do not provide people with a meaningful way to express or negotiate their preferences~\cite{frischmannRajtmajer2026}. 
It also asks for too much, as people cannot meaningfully weigh the costs and benefits of data practices across the many entities they deal with, especially as many harms only become apparent after data aggregation over time~\cite{solove2013consent}. 
User interfaces for notice and choice often contain dark patterns and condition users to agree without deliberating~\cite{frischmannSelinger2018, frischmannRajtmajer2026, habib2019optout, habib_scavenger}.

In agentic systems, the shortcomings of notice and choice are even more acute. 
To a greater extent than on the web, an agent may disclose sensitive data to downstream third parties that the user is unaware of and cannot control. 
Generally, agentic systems do not have built-in mechanisms for conveying privacy choices and minimizing data; such mechanisms are particularly important in these systems because, among other things, agents can access, use, and disclose an incredible range of sensitive data, across extended durations and contexts, and without user awareness or supervision, and can be manipulated into over-disclosing~\cite{Bagdasarian2024}.

\subsection{Privacy Preference Signals}
\label{sec:privacy-preference-signals}

The history of privacy preference signal development confirms the importance of developing the technical and legal architecture of privacy preference controls together.
On the web, the first major effort to automate notice and choice was the Platform for Privacy Preferences Project (P3P)~\cite{p3p}.
P3P allowed websites to declare their data collection practices in a machine-readable format that could be acted upon by a user agent.
However, it lacked enforcement and, thus, was not widely adopted~\cite{Cranor:Necessary:2012}.
The subsequent development of the Do Not Track (DNT) header~\cite{DoNotTrack}, which depended on self-regulation, failed for the same reason~\cite{zimmeck2026gpc}.
GPC~\cite{GPCSpec} addresses this shortcoming.
In addition to support from publishers, browser vendors, civil society organizations, and other stakeholders, it is supported by lawmakers and regulators.

In particular, the CCPA and its regulations give consumers the right to opt out of the sale and sharing of their personal information via GPC and mandate its inclusion in every browser from 2027 onward~\cite{ccpa_oops2026}.
The California, Colorado, and Connecticut attorneys general have brought GPC enforcement actions. 
Compliance is improving steadily, albeit slowly~\cite{zimmeckEtAlGPC2023, hausladenEtAlGPCWeb2025}.
GPC can also be applied to EU laws~\cite{zimmeck2026gpc}.\footnote{Advanced Data Protection Control~\cite{adpc}, another recent privacy preference signal, is specifically tailored to EU law.}
It is expressed via the header field \texttt{Sec-GPC: 1} and the JavaScript field \texttt{navigator.globalPrivacyControl} for client-side scripts. 
A site may advertise that it honors GPC by serving a \texttt{.well-known/gpc.json} resource~\cite{GPCSpec}. 
GPC alleviates the usability challenges of opting out~\cite{zimmeckEtAlGPC2024}. 

\subsection{Web Advertising and Tracking}
\label{sec:web-advertising-and-tracking}

\begin{figure}[t]
    \centering
    \begin{subfigure}[t]{0.48\textwidth}
        \centering
        \includegraphics[width=\linewidth]{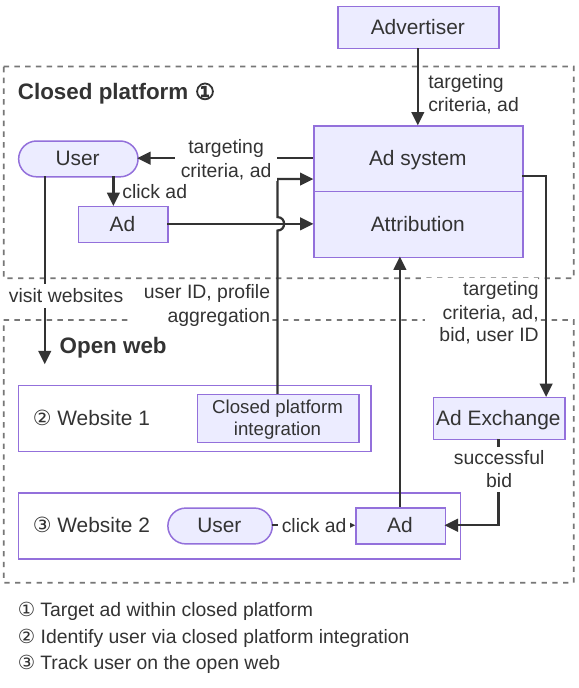}
        \caption{Open web}
        \label{fig:rtb-adtech}
    \end{subfigure}\hfill{\color{black!35}\vrule width 0.6pt}\hfill
    \begin{subfigure}[t]{0.48\textwidth}
        \centering
        \includegraphics[width=\linewidth]{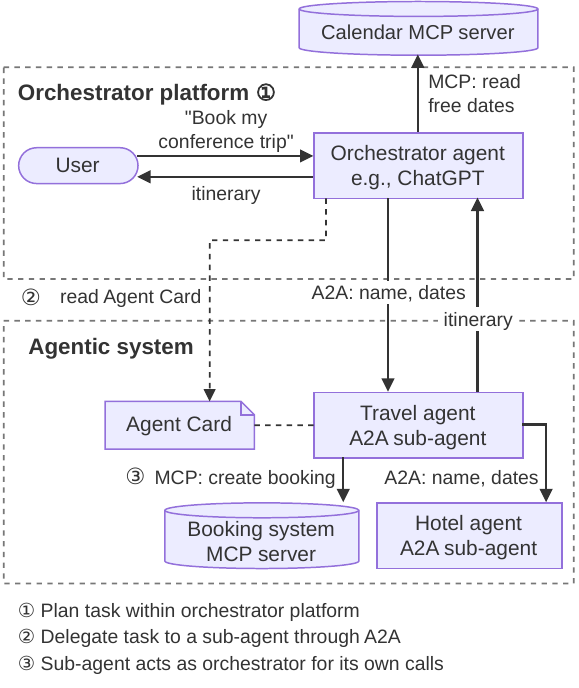}
        \caption{Agentic system}
        \label{fig:mcp-a2a}
    \end{subfigure}
    \caption{How user data travels on the open web (a) and on an agentic system (b). (a) The integration of a closed platform into the open web and its ad ecosystem. As websites integrate platform code and platforms participate in open ad exchanges, users' activities can be tracked and attributed on- and off-platform. (b)~Travel booking on an agentic system. A user asks an orchestrator agent to book a trip. The orchestrator reads the user's calendar through an MCP server, reads a travel agent's Agent Card, and delegates the booking to that agent through the A2A protocol. The travel agent then acts as an orchestrator for its own calls: it books through an MCP server and passes the user's name and dates to a hotel agent through the A2A protocol. The user's name and dates reach the travel agent, the hotel agent, and the booking system.}
    \label{fig:web-vs-agentic}
\end{figure}

Privacy preference controls matter more as agentic systems take on the tracking techniques of the web, giving third parties new ways to collect sensitive information about users and their conversations and to draw inferences from it~\cite{oliveiraEtAlButterfly2027}.
Tracking pixels, software development kits, persistent identifiers, and server-side forwarding now operate inside conversational AI services, passing prompts, generated titles, and conversation permalinks to third parties whose business models rely on advertising~\cite{oliveiraEtAlButterfly2027}.
ChatGPT serves as an example.

While OpenAI describes ad selection in ChatGPT as a vendor-internal system and states that advertisers receive neither raw conversation content nor memories~\cite{auditChatgpt1}, advertisers can install OpenAI's measurement script to send page views and purchase events to OpenAI~\cite{openaiMeasurementPixel2026}. Observations on Chrome for Android suggest that OpenAI could link this activity to ChatGPT accounts for signed-in users or device identifiers for logged-out users using \texttt{\_\_obi}, a cookie set by \texttt{bzr.openai.com} and scoped to \texttt{.openai.com}~\cite{larson2026chatgptcollector}.
This integration illustrates the broader trend: AI platforms are extending their reach beyond their internal boundaries to encompass the traditional online ad ecosystem.

Figure~\ref{fig:web-vs-agentic} contrasts how a closed platform integrates into the open web (Figure~\ref{fig:rtb-adtech}) with how an orchestrator platform integrates into an agentic system (Figure~\ref{fig:mcp-a2a}). 
For the open web, the browser is a single point at which a user’s preference, e.g., \texttt{Sec-GPC: 1}, can be attached to every outgoing request.
In an agentic system, user data traverses multiple hops of delegation, and the user typically knows only the first point of contact.
Just as social networks moved outwards over the past two decades, we expect agentic platforms to do the same.

\subsection{Privacy in Agentic Protocols}
\label{sec:privacy-in-agentic-protocols}

Agentic systems routinely delegate both tasks and authorizations to perform them~\cite{Tomasev2026}. 
Similar to the central role browsers play on the web, a central entity in agentic systems is the \emph{orchestrator}, i.e., the agent that the user interacts with and that is the originator of every outbound call made for the task the user specified. 
ChatGPT is an example of an orchestrator.\footnote{MCP calls the orchestrator the \textit{host}~\cite{ModelContextProtocolSpec}. A2A does not have a single orchestrator due to its peer-to-peer architecture.}
An orchestrator finds a remote agent through the agent's \textit{agent card}, a machine-readable description the agent publishes of what it can do and how it can be reached~\cite{A2A}.
A sub-agent receiving a call from an orchestrator can itself act as an orchestrator for any calls it makes, as shown in Figure~\ref{fig:mcp-a2a}. 

Due to this distributed nature of agentic systems, users may only know ``their'' orchestrator, and downstream agents may not know the users' identities, which gives rise to an accountability vacuum~\cite{Tomasev2026}.
However, deployers and regulators need visibility into which agent acted, and on whose behalf, to assign responsibility~\cite{Chan2024Visibility}. 
Furthermore, as model aggregators and routers, such as OpenRouter, forward requests to model providers and can inspect prompts and tool data~\cite{liu2026youragentismine}, privacy preference controls must cover these intermediaries as well.\footnote{For details see Appendix~\ref{app:model-routers}.}
Overall, security and governance in the agentic web is an open challenge~\cite{yang2025agenticweb}.

We believe that privacy preference controls and norm resolution must be built into the structure of the facilitated interactions, i.e., into the protocols that agents use to communicate and exchange data.
The most relevant protocols for agentic privacy are MCP~\cite{ModelContextProtocolSpec}, which connects agents to tools and data, and the A2A protocol, which supports agent-to-agent interaction~\cite{A2A}.\footnote{When we refer to MCP, we also include WebMCP, which connects agents to websites~\cite{WebMCPSpec}.}
MCP and A2A can serve as the foundation from which emerging ad protocols, such as the Ad
Context Protocol (AdCP)~\cite{AdCP}, Universal Commerce Protocol (UCP)~\cite{UCP}, and agentic real time bidding protocols~\cite{IABAgenticOverview}, can inherit functionality, attributes, and constraints.
While we discuss our taxonomy in the context of the MCP and A2A protocol, it is protocol-agnostic.

\subsection{Data Minimization and CI}
\label{sec:data-minimization-and-ci}

On the traditional web, the effective baseline is data maximization, which inherently limits the effectiveness of GPC and other privacy mechanisms that are pragmatically designed to operate under this baseline.
For agentic systems, we now have the opportunity to realize \textit{data minimization}, i.e., the requirement that entities collect, use, or share only necessary personal information~\cite{solove_hartzog_data_minimization}. 
The case for data minimization is simple: if data is not necessary, there is no reason to collect, use, or share it. 
While data minimization is a core principle in many privacy laws, two structural forces are blunting its effectiveness.
First, notice and choice is generally designed and implemented as a routine click-to-consent mechanism, thereby overriding data minimization laws and reinforcing data maximization~\cite{frischmannRajtmajer2026}. 
Second, data minimization rules are usually purpose-dependent, and companies tend to define purposes broadly, for example, as ``improving services.'' 
Here is where agentic systems present us with the opportunity to design and implement data minimization to user-authorized tasks rather than company-specified purposes.

Agents make minimization harder than conventional software does, because they work from the semantic context of a request rather than from structured fields alone~\cite{Bagdasarian2024, Bagdasarian2026}.
Several data minimization challenges remain~\cite{Bagdasarian2026}:
(1)~the \emph{semantic boundary problem}, determining what data a task expressed in natural language requires;
(2)~\emph{model context contamination}, ensuring minimal and appropriate information flows within an agent's reasoning cycle; and
(3)~\emph{tool over-privilege}, granting agents minimal privileges when exact data requirements are initially unknown, without burdening the user.
To address these challenges CI is a natural fit~\cite{Bagdasarian2024, Bagdasarian2026}. 
CI posits that an information flow is appropriate if it conforms to the norms of the context in which it occurs, as described by its sender, recipient, subject, information type, and transmission principle~\cite{privacy_nissenbaum_2009}. 
It has been used to detect data leakage~\cite{10.1145/3308558.3313655} and bound data disclosure~\cite{Bagdasarian2024, Bagdasarian2026}.

Various previous approaches aim to control how user data flows through agentic pipelines.
AirGapAgent uses a data minimizer to restrict an agent's access to only the user data required for the task, preventing adversarial third-party access~\cite{Bagdasarian2024}. 
Similarly, firewall architectures abstract outgoing user data and constrain incoming messages to task-specific structure~\cite{abdelnabi2025firewalls}. 
Authenticated delegation frameworks let users scope the permissions of agents acting on their behalf and propagate those limits along chains of agents~\cite{authenticated_south_2025}.
For our purposes, for the MCP implementation of context-bound data minimization, MCP Interceptors---control points that observe, filter, or modify traffic~\cite{MCPInterceptorNHIMG}---are a core building block.\footnote{The Interceptors Working Group standardizes how context operations are intercepted, validated, and transformed at key points in the agentic lifecycle~\cite{MCPInterceptorWorkingGroup}.}

\section{Methodology}
\label{sec:methodology}

We developed our taxonomy using an iterative, general taxonomy development methodology~\cite{nickerson2013method}. 
Initially, we reviewed legal, regulatory, and technical sources to derive the categories for our taxonomy (\S~\ref{sec:taxonomy_categories}), then audited the privacy preference controls for major AI products (\S~\ref{sec:audit}), and finally refined our taxonomy (\S~\ref{sec:methods-iteration}).
We also implemented prototypes for various agentic scenarios under the taxonomy, whose implementation methodology we discuss separately (\S~\ref{sec:prototypes}). 

\subsection{Taxonomy Category Derivation}
\label{sec:taxonomy_categories}

We started the derivation of our taxonomy categories with existing literature on (1) the GPC specification and the state privacy laws and regulations that give it legal effect, (2) empirical opt-out studies, and (3) studies on privacy taxonomies, rights, and consent.
The collection (\S~\ref{catB}) and secondary use, repurposing, sharing, and human review (\S~\ref{catC}) categories take into account existing privacy vocabularies and principles (e.g.,~\cite{zimmeckCloud2013, wilson2016, taxonomy_solove_2005, FTCFIPPs}).
We also analyzed work on CI (e.g., ~\cite{privacy_nissenbaum_2009}) and social boundaries (e.g., ~\cite{privacy_mulligan_2016}) for determining the context of data flows and retention limits. 
Studies on privacy by design, autonomy harms, and the engineering of human choice (e.g., ~\cite{blueprint_hartzog_2018, citronSolove2022harms, frischmannSelinger2018, frischmannOrmerod2026}) form the basis of our taxonomy categories on AI agent presence (\S~\ref{catA}), persistence (\S~\ref{catD}), and autonomy (\S~\ref{catE}).
Initially, there were disagreements between authors on how retention and data persistence should be handled by the taxonomy. 
But over multiple group meetings, we refined, reviewed, and agreed on the types and scoping decisions.

\subsection{Empirical Audit}
\label{sec:audit}

To further inform our taxonomy categories, we audited privacy preference controls in the AI products of 22 companies. 
To that end, we first identified the ten largest technology companies via three 2026 company rankings~\cite{fortuneGlobal5002026, forbesGlobal20002026, cmcTech2026} and the ten most valuable AI companies via the 2026 Forbes AI 50 ranking~\cite{forbesAI502026}.
We additionally looked at the most-used AI chatbots in Comscore's March 2026 U.S. desktop ranking (by unique visitors)~\cite{comscore2026chatbots} and added two missing companies (xAI and DeepSeek) to our list.
For each of the resulting 22 companies, we audited their AI agents and agent-building platforms available to the general public. 
Four companies had no such product, leaving 30 products from 18 companies.\footnote{See Appendix~\ref{app:audit-ppc}, Table~\ref{tab:audit-products} for the full company and product list.}

In October 2026, we collected each company's privacy policies, help pages, and developer documentation for its audited products. 
One author used an LLM (Claude Opus 5.5) to retrieve candidate pages, following the generator-verifier structure used in recent automated LLM-agent pipelines~\cite{guo2026redcodeagent, sapo}, where a retrieval agent collects the documentation unattended, while a separate verification agent independently re-checks each retrieved page. 
All retrieved pages were then manually reviewed and verified by the authors.
The authors manually added pages where the collected pages did not sufficiently document a product's settings.
Our audit involved reading these documents and then listing all documented controls that let a user, administrator, or developer stop or limit a data practice or AI behavior.

We then used the LLM to categorize the controls by what they stop or limit. 
We ran the categorization twice to check its stability; the two runs produced largely the same categories, and we reconciled the differences when reviewing them.
This process resulted in 32 candidates for preference controls emerging from the audit. 
Two authors reviewed each against our initial taxonomy and found 24 within its scope and eight outside.\footnote{See Appendix~\ref{app:audit-ppc}, Table~\ref{tab:audit-types} for all privacy preference controls found, how many products offer each, and their mappings to our taxonomy subcategories.} 
The audit also surfaced a control subcategory that our initial taxonomy did not cover, human review, which, with the agreement of all authors, we added as a subcategory to Category C (\S~\ref{catC}).
To account for the small number of products, we report each proportion with a 95\% Wilson score confidence interval.\footnote{See Figure~\ref{fig:share-by-type}.}

For each company in our audit, we also recorded the protocols it uses in its own products and the protocols that work with its models or APIs through a partner platform or a third-party package. 
Of the 22 companies, 16 use MCP and 5 use A2A in their own products.\footnote{See Appendix~\ref{app:audit-protocols}, Tables~\ref{tab:protocols-tech} and~\ref{tab:protocols-ai} for the agentic protocols that the companies use and support.} 
Counting partner platforms and third-party packages, 18 support MCP and 12 support A2A, although three A2A cases (Apple, Meta, and Databricks) rest on minimal evidence. 
MCP is the dominant protocol for connecting agents to tools and A2A for connecting agents to each other.
A2A is the only agent-to-agent protocol used by more than one company.\footnote{While Anthropic, OpenAI, xAI, and DeepSeek do not use the A2A protocol in their own products, their models work in A2A agents on partner platforms, such as Amazon Bedrock AgentCore~\cite{awsAgentCoreA2A2025} and Microsoft Foundry~\cite{msFoundryA2A}.}

We note that our audit captures the controls companies document, not whether they work as described.
Also, documentation for controls that require user login as well as controls under enterprise contracts may not have been available to us.
Further, we viewed each company's documentation and settings from a geographic location within the U.S.
Thus, we may not have seen controls available in other jurisdictions.
Finally, while we verified every retrieved page, LLM-assisted retrieval may have missed relevant documentation.

\subsection{Taxonomy Category Revision}
\label{sec:methods-iteration}

We compared each control against the initial taxonomy categories we created and revised the categories as necessary.
We checked for each control whether it was covered by at least one subcategory and mapped it accordingly.

\section{Taxonomy}
\label{sec:taxonomy}

Our taxonomy provides an overview of the different enforceable privacy preferences relevant for agentic systems.
It is based on the understanding of privacy as a right that protects an individual's personal autonomy, which not only encompasses control over data but also over environmental exposure and freedom from being influenced by others, including AI systems and agents.
It is also based on the realization of data minimization in agentic systems, resorting to signal propagation as a privacy preference control only when data disclosure is necessary to perform a task.
User-specified tasks form the contextual boundaries that AI agents must respect.
In our taxonomy, we distinguish five broad categories that capture the sequential lifecycle of agentic tasks:

\begin{enumerate}[label=\textbf{\Alph*.},ref=\Alph*]
    \item\label{tax:presence} \textit{Presence}: Whether an AI agent may be present in the user's computing environment.
    \item\label{tax:collection} \textit{Collection}: The extent of and what data the agent may collect from and about the user.
    \item\label{tax:secondary-use-category} \textit{Secondary use, repurposing, sharing, and human review}: How the agent may use collected data.
    \item\label{tax:persistence} \textit{Persistence}: Whether the agent may stay in the user’s environment after completing its task.
    \item\label{tax:autonomy} \textit{Autonomy}: Which decisions the agent can make autonomously on behalf of the user.
\end{enumerate}

\subsection{Category A: Presence}
\label{catA}

Category A covers user preferences about the presence of AI agents in their computing environment.

\begin{typology}{A}
    \item \textit{Integration preference}: AI agents may not be integrated without genuine user consent.\footnote{``Genuine user consent'' is based on demonstrably informed consent~\cite{frischmannRajtmajer2026}~\cite{frischmannVardi2025}, which requires proof that a person understands important substantive terms as a condition of consent. AI integration is substantively important. To hold otherwise would perpetuate data maximization.} AI integration must be based on the user's actual decision and not on surreptitious practices, such as automatically enabling AI features during software updates.
    \item \textit{Activation preference}: AI agents may not run in the background or as an ambient agent unless the user has affirmatively chosen so. This control applies to background and ambient agent presence; it does not apply to foreground, on-demand use. If users actively initiate an AI task, they have expressed their preference, and no separate activation preference is needed.
\end{typology}

In subcategories A1 and A2, we differentiate between consenting to AI being added and consenting to AI being active. 
Users may assert the right of each subcategory independently.
For instance, users may want AI agents to be available in their environment (A1) while not wanting them to be active (A2). 
Category A does not cover the scenario of a user directly installing agentic software.

In MCP and the A2A protocol, honoring A1 means a newly connected tool or agent, such as the travel agent in Figure~\ref{fig:mcp-a2a}, receives no user data until the user approves it~\cite{ModelContextProtocolSpec, A2A}.
Honoring A2 means A2A's long-running background tasks, which report back through push notifications while the user is away, run only if the user has turned them on~\cite{A2A}.

\subsection{Category B: Collection}
\label{catB}

Category B covers the extent of and what data the agent may collect from and about the user.

\begin{typology}{B}
    \item\label{tax:input} \textit{Input collection preference}: If a user submits data in an agentic system, for example, a prompt, text document, or video recording, the data can be stored and used to complete the task. Data unrelated to the user's task may not be collected.
    \item\label{tax:metadata} \textit{Metadata collection preference}: AI agents may observe, for example, which options users hover over on a user interface, how long they spend on a task, what text they delete before sending their prompt, or where they are currently located. Such metadata may not be collected unless it is necessary for completing the user's task.
    \begin{enumerate}[label=\textbf{B2\alph*.},ref=B2\alph*, leftmargin=*, labelindent=0pt, align=left]
        \item\label{tax:delegated-action} \textit{Delegated action collection preference}: If an agent acts on a user's behalf, it may generate a record of its work, for example, which agents it connected to and when. This record may only be created and retained to the extent it is necessary to complete the task.
    \end{enumerate}
\end{typology}

B1 and B2 are distinguished by the user's relationship to the data in question: B1 covers the data a user knowingly and actively submits, while B2 covers the data a user unknowingly produces. 
Each preference is independent.
By requesting an agent to perform a task, the user implicitly expresses the preference to input and metadata collection to the extent it is necessary to complete the task, which realizes the data minimization baseline we have adopted.

Given this baseline, by default, a user is effectively opted out of data retention for each subcategory.
The collected data may only be retained for the user's task and must be deleted once it is no longer necessary for the task, unless the law requires otherwise. 
Users can explicitly express a different retention preference, for example, that data should be stored and visible to the user beyond task completion but not be used by the system beyond the current task.

\subsection{Category C: Secondary Use, Repurposing, Sharing, and Human Review}
\label{catC}

Assuming some data collection has occurred, category C covers what agents are permitted to do with the collected data. 
By requesting an agent to perform a task, the user implicitly expresses the preference to use collected data as necessary to complete that task.
For the following subcategories, the boundary between permitted and unpermitted uses is grounded in the data minimization baseline and further determined by its context. 

\begin{typology}{C}
    \item\label{tax:secondary-use} \textit{Secondary use preferences}: An agent may not use data beyond a user-requested task unless the user explicitly expressed a preference for such use. All secondary uses, such as targeting, require users to explicitly declare their preferences.
    \begin{enumerate}[label=\textbf{C1\alph*.},ref=C1\alph*, leftmargin=*, labelindent=0pt, align=left]
        \item \textit{Targeting}: The targeting preference would allow data to be used to determine, for example, what content, options, or offers the user is shown.
        \item\label{tax:inference} \textit{Inference}: The inference preference would allow data to be used to infer, for example, attributes of the user to show personalized recommendations.
        \item \textit{Commercial and analytical use}: The commercial and analytical preference would allow, for example, data to be sold, licensed, or used for advertising, measurement, or analytics.
    \end{enumerate}
    \item\label{tax:repurposing} \textit{Data repurposing preference}: According to the data minimization default, data may not be used to train, fine-tune, improve, or evaluate AI models. Such data repurposing uses are permitted if the user explicitly expresses such a preference.
    \item\label{tax:sharing} \textit{Data sharing preference}: A single user-requested task may trigger a chain of sub-tasks across multiple connected systems or involving multiple agents (Figure~\ref{fig:web-vs-agentic}). According to the data minimization default, data may not travel further along that chain than necessary for completing the task. However, a user may express an explicit preference to allow additional sharing.
    \item \textit{Human review preference}: A user may explicitly express the preference to allow people acting for the company providing the AI agent, such as employees, contractors, or third-party reviewers, to read, listen to, or view the user's inputs, outputs, or the records of an agent's actions on behalf of the user. Legally required reviews are permitted irrespective of the user's preference.
\end{typology}

The four subcategories are distinguished by how data can be used after it is collected. 
If the user expresses respective preferences, C1 allows use for advertising, marketing, commercial, and analytical purposes, C2 allows use for AI development, C3 determines how far data travels, and C4 restricts who may see it. A user might assert the preferences for any combination of these subcategories independently.
 
\subsection{Category D: Persistence}
\label{catD}

Category D covers whether an AI agent may stay in the user's environment after completing the task the user requested it to perform.
To the extent persistence does not involve data processing or a user has already expressed the preference to allow such processing, for example, data use (category~C), it does not implicate data minimization.\footnote{It would be a worthwhile endeavor to consider the normative desirability and technical implementation of an AI minimization principle applicable to persistence of AI agents. However, this question is beyond the scope of our study here.} 
Category D has three subcategories.

\begin{typology}{D}
    \item\label{tax:post-task-presence} \textit{Post-task presence preference}: The user expresses the preference that the agent may not remain in the user's environment once it has completed the user-requested task. From a protocol perspective, upon task completion, the agent's access, integrations, scheduled jobs, and entry points are removed. Further agent activity requires invoking or installing the agent again.
    \item\label{tax:unprompted-initiation} \textit{Initiation preference}: The user expresses the preference that a persistent agent may not begin new activity without being asked. The agent may stay available, but it may not start tasks, monitor the user's environment, run scheduled or background actions, or open interactions such as reminders, suggestions, or check-ins unless the user invokes it or such activity is necessary for its persistence.
    \item\label{tax:scope-expansion} \textit{Scope expansion preference}: The user expresses the preference that a persistent agent may not extend its presence beyond its original task. The agent may stay present and may act proactively within the existing scope, but it may not enter new applications, devices, or domains, take on additional roles, or acquire new access without the user's direction.
\end{typology}

The three subcategories are distinguished by how much of the agent's presence the user allows. 
D1 prohibits any presence after the task, D2 permits presence but prohibits unprompted activity, and D3 permits presence and unprompted activity but prohibits expansion beyond the original scope. 
The subcategory preferences form a hierarchy: asserting D1 implicitly includes D2 and D3, and asserting D2 implicitly includes D3.

\subsection{Category E: Autonomy}
\label{catE}

Assuming an agent is present and active in the user's environment, category E covers which tasks an agent can perform on its own and which it must refer to the user for further instruction. 
By default, an agent may not start a task on behalf of a user.
The user may partition the tasks an agent can perform autonomously into tiers, for example, based on data sensitivity, magnitude of consequence, or reversibility of consequence.
The user may express the preference that the agent resolves tasks autonomously within some tiers while not permitting the agent to do so in others. 
When the user is not present to provide necessary instructions, the agent defaults to surfacing the decision to the user or rejecting the performance of the task. 
The agent does not have autonomy to express preferences on behalf of the user.

\subsection{Audit and Taxonomy Analysis}
\label{sec:analysis-coverage}

\begin{figure}[t]
    \centering
    \includegraphics[width=0.95\columnwidth]{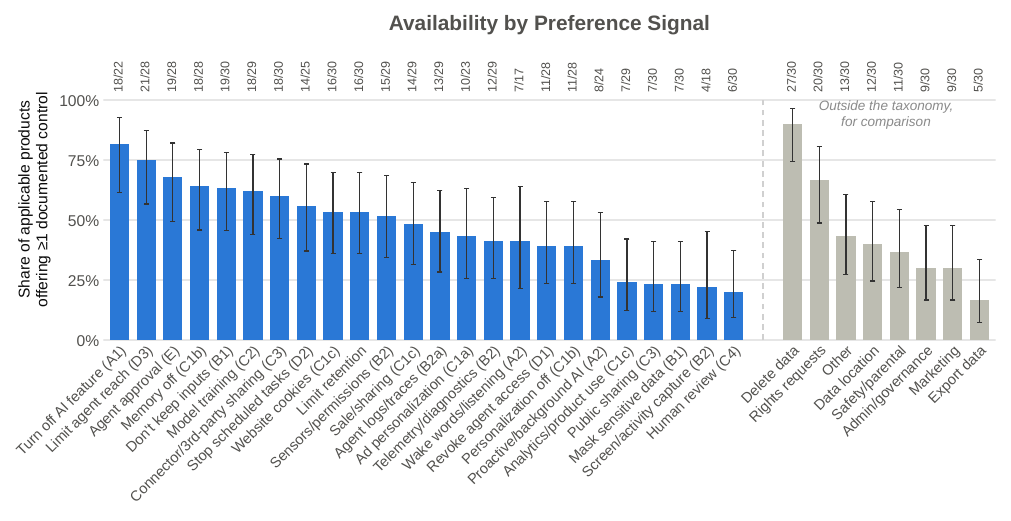}
    \caption{Share of audited AI products that offer at least one documented setting or action for a privacy preference control per a subcategory of our taxonomy  ($n = 30$ products from $k = 18$ companies).
    Denominators vary because some subcategories do not apply to every product. For instance, \textit{Turn off AI feature (A1)} does not apply to a standalone chatbot or API, where the AI is the product.
    Blue bars display the share of products covered by our taxonomy, labeled with the taxonomy subcategory they map to. Grey bars display settings and actions outside the taxonomy's scope (\S~\ref{sec:scope}), shown for comparison.
    \emph{Limit retention} has no subcategory label because it can relate to all subcategories of B (\S~\ref{catB}).
    Whiskers show 95\% Wilson score confidence intervals.}
    \label{fig:share-by-type}
\end{figure}

The results of our audit show that our taxonomy captures existing AI products' privacy preference controls well. 
Figure~\ref{fig:share-by-type} shows an overview.
Of the total 788 instances of settings and actions that we found across 30 products, 567 (72\%) fall into a subcategory, 21 (3\%) only offer preferences on how long data is kept, and the remaining 200 (25\%) fall outside the taxonomy's scope (\S~\ref{sec:scope}). 

Among all settings and actions we identified as being covered by our taxonomy, three subcategories stand out in particular. 
Only 14 of 29 products offer an opt-out of sale or sharing for advertising, including GPC (C1c).
Further, only 7 of 29 let users opt out of analytics or product-improvement use (C1c), and only 6 of 30 let users opt out of human review (C4).
These findings are particularly concerning as tracking technologies in conversational AI are proliferating~\cite{oliveiraEtAlButterfly2027}.

Our audit also shows where companies' controls are weaker than the taxonomy's definitions.
For example, input collection (B1) requires that submitted data be discarded once it is no longer necessary for a task, yet several ``temporary'' modes keep it for a set time: 30 days for Claude's incognito chats~\cite{claudeIncognito}, 72 hours for Gemini's temporary chats~\cite{geminiPrivacyHub}, and 24 hours for Perplexity's incognito threads~\cite{perplexityIncognito}.
Also, autonomy (E) requires that the agent, by default, does not start a task on behalf a user.
But Claude Code now starts interactive sessions in a mode where a classifier, rather than the user, approves actions~\cite{auditClaudeCode2}.
In each case, a control exists, but it is less restrictive than required by our taxonomy.

Even where controls exist, users often cannot tell what it does by default. 
Only 40\% of settings and actions document their default state, and the share falls to 19\% for opt-outs to be exercised by request to the company. 

\subsection{Scope and Limitations}
\label{sec:scope}
 
Our taxonomy is scoped to cover future data processing.
The deletion right, which covers past data processing, is outside its scope. 
We also exclude settings for changes to products, marketing communications, data export, plans or subscriptions, content safety, parental supervision, and labeling of AI-generated output.
Our taxonomy does not cover harms that result from an AI system's outputs. 
A system can honor the control a user exercised and still undermine their autonomy by leading or suggestive responses, for example. 
Our taxonomy is focused exclusively on realizing privacy preference controls per the CCPA, GDPR, and similar privacy and data protection laws.

Of the 200 settings and actions in our audit that fall outside the taxonomy (\S~\ref{sec:analysis-coverage}), 73 are deletion controls, which act on past rather than future processing. 
A further 34 cover privacy law requests, that is, procedural elements for exercising privacy preference controls rather than controls on their own, and 16 let users choose where data is being processed. 
The other 77 cover marketing communications, data export, content safety and parental controls, administrative settings, and other product preferences.

Finally, privacy preference controls are subject to legal requirements and terms of service, in particular, carve-outs for (1) legal obligations (e.g., OpenAI may keep deleted chats beyond its 30-day deletion window~\cite{openaiDeleteChats}), 
(2) security and abuse prevention (e.g., Google keeps all Gemini API prompts and outputs for 55 days to detect policy violations~\cite{geminiApiAbuse}), and (3) user-initiated disclosure (e.g., even with Gemini's Keep Activity turned off, a user who submits feedback also gives Google the last 24 hours of their chats, which Google may use to improve Gemini~\cite{geminiPrivacyHub}).
Thus, how much protection a privacy preference control provides ultimately depends on laws and regulations.

\section{Agentic Prototypes}
\label{sec:prototypes}

To test whether our taxonomy is feasible in practice, we built agentic prototypes that implement several of its categories and subcategories.
The eight prototypes cover different categories of our taxonomy for common agentic tasks: (1) travel planning, (2) medical questions, (3) productivity tasks, (4) website product research, (5) web search, (6) email revision, (7) general questions, and (8) travel booking.
We used MCP and the A2A protocol because our audit identified them as the dominant agentic protocols.\footnote{See Appendix~\ref{app:audit-protocols}, Tables~\ref{tab:protocols-tech} and~\ref{tab:protocols-ai}.} We propagate the GPC signal~\cite{GPCSpec} as an example in our code. In each example, the run without a preference shows how such systems commonly behave today.

\paragraph{Prototype~1: Travel Planning}
A user with privacy preferences enabled asks, ``Help me plan a 5-day trip to Japan: what should I see, eat, and know before I go?'' The assistant searches the web, synthesizes an itinerary, saves it to the user's profile, and consults what it already knows about the user before answering. The orchestrator sends the request over A2A to a search agent, which calls a \texttt{search\_web} tool over MCP.
The search agent sends the results over A2A to a synthesis agent, which writes the itinerary. The orchestrator attaches the preference to each A2A message and each MCP call, ensuring that every agent receiving the data also receives the preference (C3). The MCP server then blocks writes to the user's profile and the interaction log (B1) and blocks reads of the user's interaction history unless the user has allowed long-term profile use (C1b). Without any privacy preference, the pipeline saves the trip to the user's profile and logs the interaction. With privacy preferences set, the MCP server blocks both writes, and the pipeline keeps nothing about the trip after the conversation ends. The pipeline also stops reading the user's interaction history. The user receives the same itinerary in both cases.

\paragraph{Prototype~2: Medical Questions}
A patient asks an AI assistant, ``What does my blood pressure reading mean, and should I adjust my medication?'' The assistant retrieves the patient's medical records and answers the question. The conversation also feeds an analytics log (C1c), a model-training dataset (C2), and an external pharmaceutical advertising platform (C1a, C3). The user can turn off each secondary use separately, for example, by blocking ads while allowing analytics. Without any privacy preference, the conversation goes to all three secondary uses. With privacy preferences set, all three are turned off. The user receives the same answer in both cases.

\paragraph{Prototype~3: Productivity Tasks}
A user signs up for a productivity assistant and agrees to file access and web search. Six months later, a platform update adds two tools to the MCP server without asking the user: \texttt{email\_sender}, a communication capability, and \texttt{behavior\_tracker}, an analytics capability that runs in the background. Every tool call passes through a consent check that compares it to a versioned record of the tool categories the user agreed to. Tools in categories added after sign-up (A1) are not turned on until the user approves or declines their integration. This covers the tracker the platform starts on its own (A2, D2) and any capability beyond what the user agreed to at sign-up (D3). Without any check, both tools can be called as soon as the update ships, as if the consent given at sign-up covered them. With privacy preferences set, the new tools are turned off by default.

\paragraph{Prototype~4: Website Product Research}
A user with privacy preferences enabled asks an agent to research the iPhone 17 across eight tech publishers. The agent queries the publishers in parallel through the LLM provider's middleware, which logs every call. The preference is attached to each call sent to the publishers (C3), and each publisher decides whether to log it. The provider's middleware records which publishers the agent called, in what order, and with what subqueries (B2a). Without any privacy preference, every publisher and the provider log every call. With privacy preferences set, the publishers that honor them stop logging, but the provider's own log is basically unchanged.

\paragraph{Prototype~5: Web Search}
A user with privacy preferences enabled runs eight ordinary searches, such as metformin side effects, SNAP benefits, and signs of anxiety. None of the queries hands over personal data directly, but a classifier infers attributes about the user from each query and stores them in a profile. Before an inferred attribute is written to the profile, a check inside the search tool reads the user's preference and drops the write if the user has opted out of inference (C1b). The check is part of the tool's code. Each dropped write is logged, making the blocking visible. Without any privacy preference, the eight queries create a user profile covering health, money, housing, mental health, and job status. With privacy preferences set, the inferences are still computed but not kept. The user receives the same search results in both cases.

\paragraph{Prototype~6: Email Revision}
A user with privacy preferences enabled asks a writing assistant to polish an email to their manager about a raise. While writing, the user deletes a sentence about treatment costs, pauses for 42 seconds on the salary line, and rewrites the opening three times. The assistant keeps the draft, the editing telemetry, and what it infers from both. Each kind of data has its own blocker based on the preferences set: the draft the user submits (B1), the editing telemetry the user does not know they are creating (B2), and what the system infers from both (C1b). Without any privacy preference, all three kinds of data are kept. With privacy preferences set, all three gates block retention. The user receives the same polished email in both cases.

\paragraph{Prototype~7: General Questions}
A user talks to an assistant with memory across two sessions. In the first, they mention being vegetarian on a tight budget, and in the second, they ask about restaurants. The assistant archives each session, recalls past sessions into new ones, and builds a model of the user. The assistant supports three preference scopes: discarding each session when it ends (B1), keeping an archive the user can see but not recalling it into later sessions (C1b), or keeping transcripts but never turning them into a model of the user (C1b). Without any privacy preference, the first session is archived, recalled into the second, and used to build a behavioral model. With privacy preferences set, the first session is discarded when it ends, and nothing is recalled. The assistant still uses context from earlier in the same session to keep the conversation coherent.

\paragraph{Prototype~8: Travel Booking}
A user asks an agent to book a weekend trip, which requires the agent to search for flights, hold a hotel room, buy a non-refundable ticket, and send the user's passport details to the airline. The platform also turns on fare tracking and signs the user up for a newsletter on its own. Some of these actions cannot be undone, and some involve sensitive data. The user decides, for each type of action, whether the agent may perform it independently or must ask the user first (E). If the user has not specified a preference for an action, the agent must ask first. Fare tracking and the newsletter sign-up, which the platform starts on its own, also need the user's approval unless the user has already allowed them (A2, D2). Without any privacy preference, the platform's default tiers count as the user's consent, and four of the six actions run without the user's approval. With privacy preferences set, the platform's defaults no longer count as consent, and the agent must ask before performing any action the user has not explicitly allowed. If the user is not available to answer, the agent declines the action instead of assuming approval.

\paragraph{Lessons from the Prototypes}
By implementing these prototypes, we show that it is possible to propagate preferences in various agentic settings. These examples we show that privacy preference controls can often be implemented without significantly reducing utility. We note that there are no privacy preference-related fields in existing agentic protocols. We use the ``meta'' field in both MCP and the A2A protocol to custom-design signal propagation.
After implementing the prototypes and exploring different scenarios, we also find that the taxonomy's influence is context-dependent. The same control can cover different amounts of data and differ in whether it can be enforced or verified. Consider a user who asserts B1 (input collection) and D2 (unprompted initiation). If the agent only acts locally on the user's machine, the user can check and enforce both opt-outs. The user can inspect the device to confirm that the session data is not kept, and the agent can refer each new activity to the user as it comes up. Now suppose the same task runs on a cloud-hosted agent that completes a purchase overnight while the user is offline.
The user must trust that the company deleted the session data. 
No one is available to consult.
The opt-outs are identical in both cases, but what differs is whether anyone can confirm that they were honored.
The enforceability of preferences thus often varies significantly with context. Future work could explore ways to verify compliance in these and other scenarios.

\section{Analysis and Discussion}
\label{sec:analysis_and_discussion}

Agentic systems represent a seismic shift in computing towards substantially more autonomous activities.
However, this shift also creates the need for a broad privacy preference control landscape.

\subsection{RQ1: What privacy preference controls should people be able to express and have respected in agentic systems?}

Because AI agents act autonomously, they do more than trigger traditional privacy risks, such as unauthorized data sharing; they directly encroach on people's personal autonomy. 
As a consequence, existing privacy preference controls, such as GPC, fall short in providing the broad spectrum of controls users require in agentic systems. 
We see personal autonomy as the guiding anchor in this new environment: even as agents act on their behalf, people must remain empowered and have the rights to supervise their agents and remain in control.
Autonomy alone is too abstract a principle to design for, so our category breaks down types of controls into five categories.
Effective protection requires controls that span the entire agentic lifecycle; this includes gatekeeping initial presence and background activation, restricting data collection and secondary uses to task-bound scopes, regulating post-task persistence, and establishing authorization levels for autonomous decision-making.
We think that these are the essential controls people should have.
However, we also note that expanding the control space introduces a usability tradeoff. 
To prevent choice fatigue and avoid repeating the web's cookie banner deluge, agentic systems must focus on meaningful and effective controls.
Thus, future work in this area should include user studies to reduce cognitive burden when these controls are standardized.

\subsection{RQ2: What are the technological and legal challenges that need to be considered and resolved for privacy preference controls to be effective in agentic systems?}

In the U.S., most states still lack comprehensive privacy legislation.
There is also no legal framework to govern AI agents. 
Simply extending existing controls through new legislation and regulation will not suffice. 
Current law is structurally built around a default of data maximization, recognizing only a narrow set of controls. 
As our taxonomy demonstrates, however, agentic systems demand far broader controls that are both technically expressible via standardized protocols and legally enforceable.

Operationalizing data minimization legally represents a major challenge. 
Robust data minimization requires that (1) agents process personal data only as necessary to complete a user-specified task, and (2) any departure from this baseline requires demonstrably informed user consent~\cite{frischmannRajtmajer2026}. 
This approach evaluates necessity against bounded and user-authorized tasks rather than open-ended business-defined purposes. 
However, establishing clear task boundaries is challenging. 

In this regard, context, in particular, CI, has emerged as a promising boundary construct~\cite{privacy_nissenbaum_2009, Bagdasarian2026}.
However, mapping legal and social concepts of context onto technical systems is a major challenge. 
While technical mechanisms, such as MCP Interceptors, can enforce discrete operational boundaries, for example, isolating chat sessions, it is unclear whether those align with psychological or social norms and realities. 
From a legal perspective, a major challenge lies in shifting statutory baselines from purpose-based data maximization to task-bounded minimization, while the principal technological challenge requires guaranteeing end-to-end signal verifiability and auditability across opaque, multi-hop delegation chains.
Bringing these disparate strands together will be critical for meaningful and impactful privacy protection on the agentic web.

\subsection{RQ3: How can we structurally integrate privacy preference controls into agentic systems, especially as such systems interact with the web and online ad ecosystem?}

AI agents have the potential to act as active privacy shields for users~\cite{zimmeck2026hill}. 
Realizing this potential, however, requires embedding controls directly into the structure of agentic protocols.
Our prototypes demonstrate that such integration is feasible, shifting the challenge toward industry standardization.
Privacy preference controls and natural task boundaries must be woven into the architecture of agentic systems. 
Structurally, this can be achieved by embedding mandatory preference---checking interceptors and standardized header signals---to ensure that data minimization defaults are enforced at every delegation point in a pipeline, especially, before agents interact with web endpoints or ad networks.
\section{Conclusion}
\label{sec:conclusions}

In this study, we systematized what enforceable privacy preferences a user can and should be able to exercise in agentic systems. 
If we are to move away from the data maximization defaults of our current ecosystem, we require an integrated techno-legal effort to shape the legal system and the build-out of the technological infrastructure of agentic systems. 
The varied stakeholders in the agentic ecosystem include a vast and opaque network of agents, sub-agents, and third parties, all of whom need to achieve consensus to appropriately protect privacy. 
Developing both new privacy-preserving technology and mapping existing legal concepts, such as controllers/processors~\cite{ccpa-cpra, gdpr}, onto the components of agentic systems remains a rich area for future work and is necessary to ensure the agentic ecosystem can be built with privacy in mind from the ground up.
We see this juncture as an opportunity to build the right privacy primitives while the agentic web is still in its formative stage to structurally protect people's privacy and personal autonomy.

\bibliographystyle{ACM-Reference-Format}
\bibliography{references.bib}

\clearpage
\appendix
\raggedbottom
\setlength{\LTpre}{\medskipamount}
\setlength{\LTpost}{\medskipamount}
\captionsetup[longtable]{position=bottom,skip=6pt}

\section{Audit of Privacy Preference Controls}
\label{app:audit-ppc}

Table~\ref{tab:audit-types} lists the types of opt-out we found in our audit and the taxonomy subcategory each maps to. Table~\ref{tab:audit-products} lists the companies and products we audited, with one representative source page per product. To audit each product, we also followed many other pages and paths through the product itself, such as settings screens, help pages, and policies.
\begingroup
\footnotesize
\begin{longtable}{@{}l >{\raggedright\arraybackslash}p{0.48\columnwidth} l r r@{}}
\toprule
\textbf{Code} & \textbf{Type of opt-out} & \textbf{Taxonomy} & \textbf{Products} & \textbf{Controls} \\
\midrule
\endfirsthead
\toprule
\textbf{Code} & \textbf{Type of opt-out} & \textbf{Taxonomy} & \textbf{Products} & \textbf{Controls} \\
\midrule
\endhead
\midrule
\multicolumn{5}{r}{\textit{Continued on next page}} \\
\endfoot
\endlastfoot
T01 & Turn an AI feature or assistant on or off & A1 & 18 of 22 & 67 \\
T02 & Stop wake words and always-on listening & A2 & 7 of 17 & 8 \\
T03 & Stop proactive suggestions and background AI & A2 & 8 of 24 & 16 \\
T04 & Do not keep what I submit & B1 & 19 of 30 & 35 \\
T05 & Filter or mask sensitive data & B1 & 7 of 30 & 9 \\
T06 & Limit device sensors and permissions & B2 & 15 of 29 & 22 \\
T07 & Limit usage telemetry and diagnostics & B2 & 12 of 29 & 28 \\
T08 & Limit screen and activity capture & B2 & 4 of 18 & 9 \\
T09 & Limit logs and traces of agent and model activity & B2a & 13 of 29 & 26 \\
T10 & Set how long data is kept & Retention note (B) & 16 of 30 & 21 \\
T11 & Opt out of ad personalization & C1a & 10 of 23 & 18 \\
T12 & Turn off or limit memory & C1b & 18 of 28 & 32 \\
T13 & Turn off personalization from other data & C1b & 11 of 28 & 23 \\
T14 & Opt out of sale or sharing for ads, incl. GPC & C1c & 14 of 29 & 21 \\
T15 & Refuse non-essential cookies on vendor websites & C1c & 16 of 30 & 22 \\
T16 & Opt out of analytics, feedback and product improvement & C1c & 7 of 29 & 10 \\
T17 & Opt out of model training & C2 & 18 of 29 & 37 \\
T18 & Control which connectors and third parties receive data & C3 & 18 of 30 & 39 \\
T19 & Control public sharing of chats and outputs & C3 & 7 of 30 & 10 \\
T20 & Opt out of human review & C4 & 6 of 30 & 9 \\
T21 & Revoke an agent's access and clear its working state & D1 & 11 of 28 & 12 \\
T22 & Pause or delete scheduled and background agent tasks & D2 & 14 of 25 & 18 \\
T23 & Limit which apps, sites, files and networks an agent can reach & D3 & 21 of 28 & 41 \\
T24 & Require approval before agent actions & E & 19 of 28 & 55 \\
\midrule
T25 & Choose where data is processed or stored & Application of taxonomy & 12 of 30 & 16 \\
T26 & Delete data & Out of scope & 27 of 30 & 73 \\
T27 & Request privacy rights from the vendor & Legal mechanism & 20 of 30 & 34 \\
T28 & Opt out of marketing communications & Out of scope & 9 of 30 & 13 \\
T29 & Export data & Out of scope & 5 of 30 & 5 \\
T30 & Content safety, parental and age controls & Out of scope & 11 of 30 & 14 \\
T31 & Admin access, security and release governance & Out of scope & 9 of 30 & 26 \\
T32 & Other & Out of scope & 13 of 30 & 19 \\
\bottomrule
\caption{Types of privacy preference controls found in our audit of 30 AI products, and the taxonomy subcategory each maps to. \emph{Products} counts products with at least one verified setting or action of the type, out of the products where the type applies. \emph{Controls} counts verified settings and actions. T25 is an application of the taxonomy; T27 is a legal mechanism to implement parts of the taxonomy; T26 and t28-T32 are out of scope for the taxonomy (\S~\ref{sec:scope}).}
\label{tab:audit-types}\\
\end{longtable}
\endgroup

\begingroup
\footnotesize
\begin{longtable}{@{}l l l r l@{}}
\toprule
\textbf{Company} & \textbf{Sample} & \textbf{Product} & \textbf{Controls} & \textbf{Sources} \\
\midrule
\endfirsthead
\toprule
\textbf{Company} & \textbf{Sample} & \textbf{Product} & \textbf{Controls} & \textbf{Sources} \\
\midrule
\endhead
\midrule
\multicolumn{5}{r}{\textit{Continued on next page}} \\
\endfoot
\endlastfoot
Apple & Top 10 tech & Apple Intelligence / Siri & 32 & \cite{auditAppleIntelligenceSiri1} \\
Alphabet (Google) & Top 10 tech; Comscore \#2 & Gemini (app) & 27 & \cite{auditGeminiApp1} \\*
 &  & Gemini in Workspace and Chrome & 19 & \cite{auditGeminiInWorkspaceAndCh1} \\*
 &  & Gemini API / Vertex AI & 18 & \cite{auditGeminiApiVertexAi1} \\
Microsoft & Top 10 tech; Comscore \#3 & Microsoft Copilot (app) & 49 & \cite{auditMicrosoftCopilotApp1} \\*
 &  & Copilot in Microsoft 365 / Windows & 28 & \cite{auditCopilotInMicrosoft365W1} \\*
 &  & Microsoft Foundry & 11 & \cite{auditMicrosoftFoundry1} \\
Nvidia & Top 10 tech & NVIDIA G-Assist & 17 & \cite{auditNvidiaGAssist1} \\*
 &  & NVIDIA NIM / NeMo Agent Toolkit & 20 & \cite{auditNvidiaNimNemoAgentTool1} \\
Meta & Top 10 tech & Meta AI & 31 & \cite{auditMetaAi1} \\
TSMC & Top 10 tech & \emph{No AI assistant or agent product} & -- & -- \\
Samsung Electronics & Top 10 tech & Samsung Galaxy AI / Bixby & 39 & \cite{auditSamsungGalaxyAiBixby1} \\
SK Hynix & Top 10 tech & \emph{Agent framework is internal only} & -- & -- \\
Tencent & Top 10 tech & Tencent Yuanbao & 24 & \cite{auditTencentYuanbao1} \\*
 &  & Tencent Cloud ADP & 15 & \cite{auditTencentCloudAdp1} \\
Broadcom & Top 10 tech & Broadcom Tanzu AI & 36 & \cite{auditBroadcomTanzuAi1, auditBroadcomTanzuAi2} \\
Anthropic & Top 10 AI; Comscore \#4 & Claude (apps) & 38 & \cite{auditClaudeApps1} \\*
 &  & Claude API & 17 & \cite{auditClaudeApi1} \\*
 &  & Claude Code & 35 & \cite{auditClaudeCode1} \\
OpenAI & Top 10 AI; Comscore \#1 & ChatGPT & 48 & \cite{auditChatgpt1} \\*
 &  & dots (OpenAI) & 16 & \cite{auditDotsOpenai1} \\*
 &  & OpenAI API & 21 & \cite{auditOpenaiApi1} \\
Databricks & Top 10 AI & Databricks Agent Bricks & 51 & \cite{auditDatabricksAgentBricks1} \\
Safe Superintelligence & Top 10 AI & \emph{No product to audit} & -- & -- \\
Perplexity & Top 10 AI; Comscore \#7 & Perplexity & 26 & \cite{auditPerplexity1} \\*
 &  & Comet (Perplexity) & 20 & \cite{auditCometPerplexity1} \\
Sierra & Top 10 AI & Sierra agent platform & 33 & \cite{auditSierraAgentPlatform1} \\
Applied Intuition & Top 10 AI & Applied Intuition Dana & 15 & \cite{auditAppliedIntuitionDana1} \\
Skild AI & Top 10 AI & \emph{No product to audit} & -- & -- \\
Mistral AI & Top 10 AI & Le Chat (Mistral) & 25 & \cite{auditLeChatMistral1} \\
Thinking Machines Lab & Top 10 AI & Thinking Machines Tinker & 25 & \cite{auditThinkingMachinesTinker1} \\
xAI & Comscore \#5 & Grok & 22 & \cite{auditGrok1} \\*
 &  & Grok Bot & 21 & \cite{auditGrokBot1} \\
DeepSeek & Comscore \#6 & DeepSeek & 9 & \cite{auditDeepseek1} \\
\bottomrule
\caption{Companies and products in our audit. \emph{Sample} gives the list that put the company in the sample. \emph{Controls} counts verified settings and actions. \emph{Sources} cites one representative page per product.}
\label{tab:audit-products}\\
\end{longtable}
\endgroup

\section{Audit of Protocols}
\label{app:audit-protocols}

Table~\ref{tab:protocols-tech} lists the protocols used by top tech companies. Table~\ref{tab:protocols-ai} lists protocols used by top AI companies. Each company cites one representative documentation page; each entry also draws on other pages of the company's documentation, specifications, code repositories, and partner platforms.

\begingroup
\footnotesize
\begin{longtable}{@{}>{\raggedright\arraybackslash}p{0.16\textwidth} >{\raggedright\arraybackslash}p{0.36\textwidth} >{\raggedright\arraybackslash}p{0.25\textwidth} >{\raggedright\arraybackslash}p{0.15\textwidth}@{}}
\toprule
\textbf{Company} & \textbf{Used internally} & \textbf{Supported} & \textbf{Own framework} \\
\midrule
\endfirsthead
\toprule
\textbf{Company} & \textbf{Used internally} & \textbf{Supported} & \textbf{Own framework} \\
\midrule
\endhead
\midrule
\multicolumn{4}{r}{\textit{Continued on next page}} \\
\endfoot
\endlastfoot
Apple~\cite{appleXcodeMCP} & MCP; opposes WebMCP (WebKit position) & MCP, A2A & App Intents \\
Alphabet (Google)~\cite{googleMCP} & MCP, A2A (created), UCP and WebMCP (co-created), AP2 (created), AG-UI, Web Bot Auth (testing) & See under ``Used internally'' & ADK \\
Microsoft~\cite{msAgentFramework} & MCP, A2A, WebMCP (co-created), NLWeb (created), AG-UI, Agentic Commerce Protocol, UCP & See under ``Used internally'' & Agent Framework \\
Nvidia~\cite{nvidiaNeMoAgentToolkit} & MCP, A2A & See under ``Used internally'' & NeMo Agent Toolkit \\
Meta~\cite{metaLlamaStack} & MCP; UCP (council member)
 & MCP, A2A & None found \\
TSMC & None found & None found & None \\
Samsung Electronics~\cite{samsungResearch6GAgents} & MCP and A2A
 & MCP & None found
 \\
SK Hynix & None found
 & None found \\
Tencent~\cite{tencentADP} & MCP, AG-UI & MCP, AG-UI, A2A & ADP \\
Broadcom~\cite{tanzuMCP} & MCP, A2A & See under ``Used internally'' & Tanzu Platform \\
\bottomrule
\caption{Agentic protocols used and supported by the ten largest technology companies.}
\label{tab:protocols-tech}\\
\end{longtable}
\endgroup

\begingroup
\footnotesize
\begin{longtable}{@{}>{\raggedright\arraybackslash}p{0.16\textwidth} >{\raggedright\arraybackslash}p{0.36\textwidth} >{\raggedright\arraybackslash}p{0.25\textwidth} >{\raggedright\arraybackslash}p{0.15\textwidth}@{}}
\toprule
\textbf{Company} & \textbf{Used internally} & \textbf{Supported} & \textbf{Own framework} \\
\midrule
\endfirsthead
\toprule
\textbf{Company} & \textbf{Used internally} & \textbf{Supported} & \textbf{Own framework} \\
\midrule
\endhead
\midrule
\multicolumn{4}{r}{\textit{Continued on next page}} \\
\endfoot
\endlastfoot
Anthropic~\cite{anthropicMCP} & MCP (created), MCP Apps (co-created), Agent Skills (created), AGENTS.md & See under ``Used internally'', plus A2A and AG-UI & Claude Agent SDK \\
OpenAI~\cite{openaiAgentsSDKMCP} & MCP, Agentic Commerce Protocol (co-created), AGENTS.md (created), MCP Apps (co-created), WebMCP, Web Bot Auth & See under ``Used internally'', plus A2A & Agents SDK \\
Databricks~\cite{databricksMCP} & MCP, A2A & See under ``Used internally'' & Agent Bricks \\
Safe Superintelligence & None found & None found & No product \\
Perplexity~\cite{perplexityAgentMCP} & MCP & See under ``Used internally'' & Agent API \\
Sierra~\cite{sierraMCPGateway} & MCP & See under ``Used internally'' & Agent SDK \\
Applied Intuition~\cite{appliedProducts} & MCP & See under ``Used internally'' & Dana \\
Mistral AI~\cite{mistralVibeConfig} & MCP, AGENTS.md & See under ``Used internally'' & Agents API \\
Thinking Machines Lab~\cite{tinkerDocs} & None found & MCP & Tinker API \\
xAI (Grok)~\cite{xaiRemoteMCP} & MCP, AGENTS.md
 & See under ``Used internally'', plus A2A & xAI SDK, Grok Build \\
DeepSeek~\cite{deepseekmcpclient2026} & MCP, ACP, AGENTS.md, Agent Skills & See under ``Used internally'', plus A2A & DeepSeek Harness \\
\bottomrule
\caption{Agentic protocols used and supported by the ten Forbes AI 50 (2026) companies with the highest valuation from a completed funding round (above the rule) and by the developers of the two additional chatbots from Comscore's March 2026 U.S.\ desktop ranking (below the rule). }
\label{tab:protocols-ai}\\
\end{longtable}
\endgroup

\section{Prototypes}
\label{app:prototypes}

Table~\ref{tab:opt-out-typology} summarizes our taxonomy's opt-out categories and subcategories, and maps each category to the prototypes in our repository that demonstrate it; each prototype can be run to see what the categories look like in real-world examples.

\begingroup
\footnotesize
\setlength{\tabcolsep}{4pt}
\begin{longtable}{@{}>{\raggedright\arraybackslash}p{\dimexpr0.16\textwidth-6pt\relax} >{\raggedright\arraybackslash}p{\dimexpr0.26\textwidth-6pt\relax} >{\raggedright\arraybackslash}p{\dimexpr0.47\textwidth-6pt\relax} >{\raggedright\arraybackslash}p{\dimexpr0.11\textwidth-6pt\relax}@{}}
\toprule
\textbf{Category} & \textbf{Subcategory} & \textbf{Governs} & \textbf{Prototype} \\
\midrule
\endfirsthead
\toprule
\textbf{Category} & \textbf{Subcategory} & \textbf{Governs} & \textbf{Prototype} \\
\midrule
\endhead
\midrule
\multicolumn{4}{r}{\textit{Continued on next page}} \\
\endfoot
\endlastfoot
\multirow[t]{2}{*}{A. Presence}
 & A1. Integration & Whether AI can be added to a product without genuine user consent & 3 \\*
 & A2. Activation & Whether AI can run ambiently or passively without affirmative choice & 3, 8 \\
\midrule
\multirow[t]{4}{*}{B. Collection}
 & B1. Input & What the user knowingly submits, beyond what the task needs & 1, 6, 7 \\*
 & B2. Metadata & What the user passively generates while interacting & 6 \\*
 & \quad B2a. Delegated action & What an agent records about the actions it takes on the user's behalf & 4 \\*
 & \quad Retention (note) & How long collected data is kept after the task; by default, not beyond it & 7 \\
\midrule
\begin{tabular}[t]{@{}l@{}}C. Secondary use, \end{tabular}
 & C1. Secondary use & Any use of data beyond the user-requested task & -- \\*
 repurposing, sharing, & \quad C1a. Targeting & Whether data determines what content, options, or offers the user sees & 2 \\*
 and human review & \quad C1b. Inference & Whether data tailors responses, tone, or recommendations based on inferred preferences or past behavior & 1, 5, 6, 7 \\*
 & \quad C1c. Commercial and analytical & Whether data is sold, licensed, or used for advertising, measurement, or analytics & 2 \\*
 & C2. Data repurposing & Whether data is used to train, fine-tune, improve, or evaluate AI models & 2 \\*
 & C3. Sharing & How far data travels along chains of connected systems or agents & 1, 2, 4 \\*
 & C4. Human review & Whether people acting for the provider may read, listen to, or view the user's data or the agent's records & -- \\
\midrule
\multirow[t]{3}{*}{D. Persistence}
 & D1. Post-task presence & Whether the agent remains in the user's environment after its task is complete & -- \\*
 & D2. Unprompted initiation & Whether a persistent agent may begin new activity without being invoked & 3, 8 \\*
 & D3. Scope expansion & Whether a persistent agent may extend its presence beyond its original task & 3 \\
\midrule
E. Autonomy & -- & How much discretion the agent has to resolve decisions without referring them back to the user & 8 \\
\bottomrule
\caption{Summary of taxonomy categories, subcategories, and scenarios they govern in the agentic web.}
\label{tab:opt-out-typology}\\
\end{longtable}
\endgroup

\section{Implementation Details}
\label{app:technical-analysis}

Table~\ref{tab:implementation-tradeoffs} compares concrete implementations of the categories in Section~\ref{sec:taxonomy}. The examples assume an established preference and permitted task scope in the search, medical, memory, and booking scenarios in Section~\ref{sec:prototypes}. They are recommendations, not new taxonomy categories or measured comparisons of attack success and latency. Retention follows Category~\ref{tax:collection}'s task-based rule (Section~\ref{catB}), including explicit retention choices and legally required exceptions. This prospective rule does not provide a retrospective deletion right. Category~\ref{tax:persistence} instead governs the agent's post-task presence (\ref{tax:post-task-presence}), unprompted initiation (\ref{tax:unprompted-initiation}), and scope expansion (\ref{tax:scope-expansion}) (Section~\ref{catD}).

\begingroup
\footnotesize
\renewcommand{\arraystretch}{1.15}\setlength{\tabcolsep}{4pt}
\begin{longtable}{@{}>{\raggedright\arraybackslash}p{0.22\textwidth} >{\raggedright\arraybackslash}p{0.37\textwidth} >{\raggedright\arraybackslash}p{0.37\textwidth}@{}}
\toprule
\textbf{Existing preference} & \textbf{Implementation choice} & \textbf{When to use it and what can fail} \\
\midrule
\endfirsthead
\toprule
\textbf{Existing preference} & \textbf{Implementation choice} & \textbf{When to use it and what can fail} \\
\midrule
\endhead
\midrule
\multicolumn{3}{r}{\textit{Continued on next page}} \\
\endfoot
\endlastfoot
No unnecessary metadata collection (\ref{tax:metadata})
& Disable listeners that record deleted text or pause duration unless those signals are needed for the task. Filtering at an analytics collector is a later check.
& Prefer the source check when capture itself is prohibited. Collector filtering fits an existing pipeline but may leave copies in client buffers or request logs. \\
Secondary use, model training, and sharing (\ref{tax:secondary-use}, \ref{tax:repurposing}, \ref{tax:sharing})
& Check calls before sending them. Add a gateway for shared destinations and recipient checks before profile writes or training-job publication.
& Use a gateway to share rules across agents, with direct network bypass blocked. Recipient checks are needed for internal processing; a forwarded preference or acknowledgment cannot establish compliance. \\
No inference beyond the task (\ref{tax:inference})
& Block a health-inference classifier unrelated to the user's task before it runs. Blocking only its profile write prevents storage, not inference.
& Suitable for a known inference job. A general-purpose model may still make unrelated inferences while answering a query, which limits the guarantee. \\
Task-limited retention (\ref{tax:input}, \ref{tax:metadata}, \ref{tax:delegated-action})
& Keep task state in working memory and discard it when no longer needed. Alternatives delete stored records or erase a separate task encryption key.
& Memory limits crash recovery; Redis snapshots and write logs must also be disabled.\footnotemark DynamoDB expiry may leave records for days.\footnotemark Key erasure requires eliminating recoverable key copies and does not physically remove ciphertext~\cite{nistMediaSanitization2025}. \\
User-visible history without secondary use (\ref{tax:collection}, \ref{tax:secondary-use}, \ref{tax:inference})
& Where the user permits retention, give the history viewer and agent separate permissions. Check source-task eligibility and block unauthorized profile builders and summary writes.
& Storage checks cover more callers than query filters. PostgreSQL's row security restricts accessible records, but privileged roles can bypass it.\footnotemark Cached summaries and vector indexes must preserve the restrictions. \\
Selective delegation (\ref{tax:autonomy})
& An executor matches the final request to an approval record and admits the operation once. Retry with the same provider-recognized identifier or reconcile the result.
& Needed for purchases, sends, and sensitive disclosures. Stripe may remove deduplication keys after at least 24 hours;\footnotemark an unknown result after that window cannot justify a fresh purchase. \\
\bottomrule
\caption{Choices for enforcing existing privacy preferences. The recommended check depends on the required outcome. Preventing storage does not necessarily prevent collection or inference.}
\label{tab:implementation-tradeoffs}\\
\end{longtable}
\endgroup
\addtocounter{footnote}{-3}\footnotetext{\url{https://redis.io/docs/latest/operate/oss_and_stack/management/persistence/}.}
\stepcounter{footnote}\footnotetext{\url{https://docs.aws.amazon.com/amazondynamodb/latest/developerguide/TTL.html}.}
\stepcounter{footnote}\footnotetext{\url{https://www.postgresql.org/docs/17/ddl-rowsecurity.html}.}
\stepcounter{footnote}\footnotetext{\url{https://docs.stripe.com/api/idempotent_requests}.}

Two operational checks apply across these examples. A training export queued before a user withdraws permission must check the current preference when it is admitted; holding the export during a preference-store outage preserves the restriction but delays completion. An access log containing the full request body can retain a conversation after the application deletes it. Even a log limited to an operation identifier and decision describes delegated activity, which the delegated-action collection privacy preference (\ref{tax:delegated-action}) restricts, and must meet the chosen retention limit.

\begin{figure}[!htbp]
\centering
\includegraphics[scale=0.77]{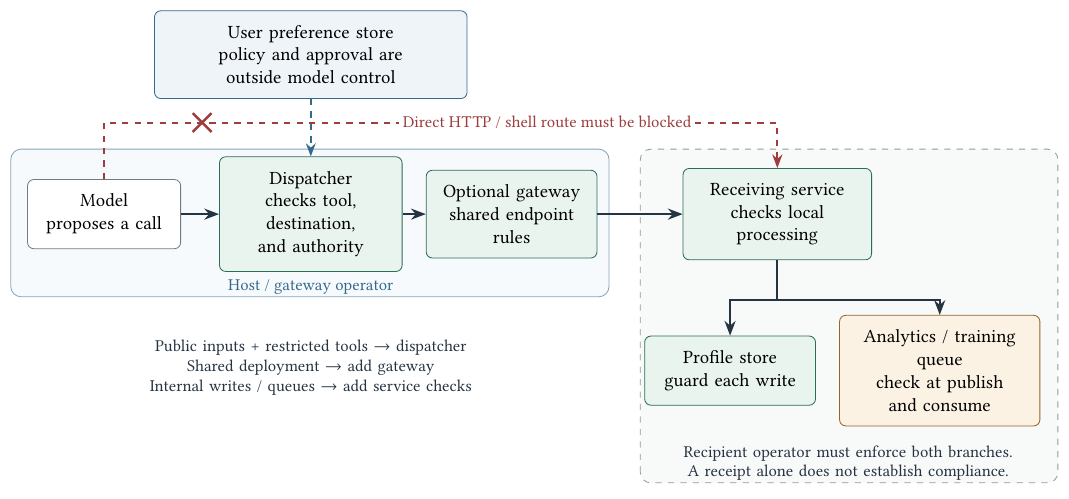}
\caption{Where to enforce a preference. Application code checks proposed calls, an optional gateway shares destination rules, and cooperating recipient services check internal writes and secondary jobs. An unrestricted direct network route bypasses the dispatcher. Enforcement at an independent recipient still depends on its operator.}
\Description{User preferences feed a dispatcher that checks model-proposed calls. Requests pass through an optional gateway to a recipient with separate profile and secondary-job checks. A crossed-out direct network route shows the bypass that must be prevented.}
\label{fig:technical-enforcement}
\end{figure}

\begin{figure}[!htbp]
\centering
\includegraphics[scale=0.77]{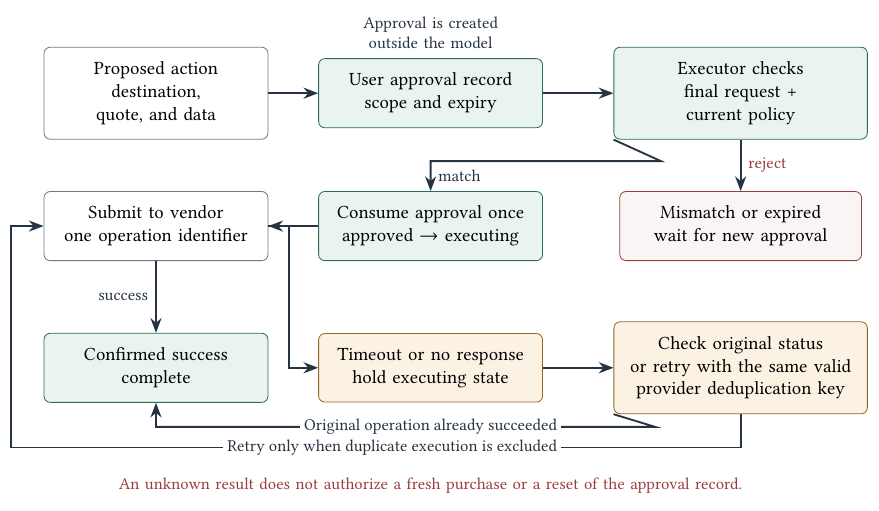}
\caption{Approval for a consequential action. The executor checks the final request against approval created outside the model and checks the current preference. A mismatch needs new approval. An uncertain result requires a retry protected against duplicate execution or a check of the original operation's status.}
\Description{A proposal passes through user approval and an executor check. Matching requests consume approval and reach the vendor; mismatches wait. Success ends execution, while timeouts remain unresolved until the original status is checked or a valid duplicate-prevention mechanism permits retry.}
\label{fig:technical-approval}
\end{figure}

Figure~\ref{fig:technical-enforcement} shows where these checks belong. Figure~\ref{fig:technical-approval} details approval and retry handling for consequential actions.

\subsection{Privacy at Intermediaries}
\label{app:model-routers}

Intermediaries such as routers can select cheaper or faster providers and redirect requests when a provider is unavailable.\footnote{\url{https://openrouter.ai/docs/guides/routing/provider-selection}.} Each added service, however, gives another operator access to prompts and tool data and creates another place where that information can be retained or stolen~\cite{liu2026youragentismine}. The final provider's promise not to retain prompts does not cover a relay's own logs. OpenRouter, for example, describes prompt and response retention as opt-in while retaining request metadata.\footnote{\url{https://openrouter.ai/docs/guides/privacy/data-collection}.} Users can separately exclude upstream providers that train on inputs or require endpoints that do not retain data.\footnote{\url{https://openrouter.ai/docs/guides/features/zdr}.} We did not test whether providers honor these settings.

These requirements become harder to check when the recipients are not fully known. Among 428 paid and free routers, mostly sourced from Chinese-language marketplaces and communities, Liu et al.\ found nine injecting malicious code and 17 accessing researcher-owned cloud credentials~\cite{liu2026youragentismine}. Ji et al.'s measurements across 39 reseller endpoints revealed shared access to upstream prompt caches, evidence of hidden dependencies rather than verified supplier relationships~\cite{ji2026cachetracer}. Chen examines how routing metadata can expose sensitive topics even when prompt logging is disabled~\cite{chen2026routingmetadata}. Checking only the final provider would leave intermediate storage, reuse, and disclosure outside the check. We therefore recommend excluding routes whose intermediaries' declared data practices conflict with the user's preferences, even when this limits the available services or delays the task.

\FloatBarrier
\Needspace{10\baselineskip}
\section{Prototype Scenarios}
\label{app:prototype-scenarios}

\begingroup
\footnotesize
\renewcommand{\arraystretch}{1.15}\setlength{\tabcolsep}{4pt}
\begin{longtable}{@{}c >{\raggedright\arraybackslash}p{0.465\textwidth} >{\raggedright\arraybackslash}p{0.465\textwidth}@{}}
\toprule
\textbf{\#} & \textbf{Scenario} & \textbf{What it demonstrates} \\
\midrule
\endfirsthead
\toprule
\textbf{\#} & \textbf{Scenario} & \textbf{What it demonstrates} \\
\midrule
\endhead
\midrule
\multicolumn{3}{r}{\textit{Continued on next page}} \\
\endfoot
\endlastfoot
1
  & An AI agent plans a five-day trip to Japan, saves the itinerary to the user's profile, and shares it with a company.
  & Tool-level blocking: The opt-out signal travels with every MCP tool call. Specified tools are blocked from storing and handling data. \\
2
  & A patient uses a medical assistant. The conversation feeds data for different purposes, such as analytics, model training, and ad targeting.
  & Gating after the agent: Pipeline supports different opt-outs per purpose. \\
3
  & An app user consents to file access and web search initially. A later update silently adds an email sender and a behavior tracker.
  & Consent-scoped tools: Tools added after consent was given receive no data from the pipeline by default. \\
4
  & A user asks an assistant to research the iPhone 17 across eight tech publishers at once.
  & Signal propagation: The signal can be forwarded to each publisher, and publishers that honor it stop tracking. The provider layer in between still logs every query, so its view does not shrink; recording whether the signal was set turns the opt-out itself into tracked data; and if any hop drops the signal, no downstream site can enforce the opt-out. \\
5
  & A user runs eight ordinary searches, and a classifier infers health, financial, and employment details the user never disclosed.
  & Inference firewall: Opting out of inference blocks derived attributes from being written to the user's profile, while the searches themselves still run. \\
6
  & A user ask an AI agent to polish an email to their manager about a raise, deleting a sensitive line, pausing on the salary figure, and rewriting the opening several times.
  & Collection: Collection can be limited at three levels: submitted input is not logged, behavioral traces (edits, pauses, deletions) are not recorded, and inferences are blocked and labeled by source. \\
7
  & A user tells a memory-enabled assistant they are vegetarian on a tight budget, then returns in a second session.
  & Retention: No data retention means context is dropped at session end, no recall across sessions, and no long-term profile. Context within the current session is preserved. \\
8
  & A travel agent books a weekend trip involving six actions of differing risk, from searching flights to purchasing a non-refundable ticket and sharing passport data.
  & Autonomy: Users assign actions to autonomy tiers; user choices override vendor defaults, and unassigned actions default to the strictest tier, so purchases, sharing of sensitive data, tracking, and subscriptions wait for explicit approval. \\
\bottomrule
\caption{Prototype implementations and the privacy preferences each demonstrates.}
\label{tab:def-prototypes}\\
\end{longtable}
\endgroup

\end{document}